\documentclass[]{spie}  

\usepackage{amsmath,amsfonts,amssymb}
\usepackage{graphicx}
\usepackage{textgreek}
\usepackage[colorlinks=true, allcolors=blue]{hyperref}

\title{PRIMA prototype spaceflight KID readout electronics laboratory characterization}

\author[1,5]{Adrian K. Sinclair}
\author[2,6]{Chris Albert}
\author[2]{C. Matt Bradford} 
\author[1]{Patrick Brown} %
\author[3]{Sean Bryan} %
\author[1]{Jesse Coldsmith} %
\author[1]{Jennifer Corekin}
\author[1]{Sumit Dahal}
\author[1]{Thomas K. Devlin}
\author[1]{Thomas Essinger-Hileman}
\author[2,6]{Logan Foote}
\author[2]{Marc Foote} 
\author[1]{Draisy Friedman}
\author[1]{Alessandro Geist}
\author[1]{Jason Glenn}
\author[1]{Christopher Green}
\author[1]{Kyle Helson}
\author[1]{Kevin Horgan}
\author[3]{Tracee Jamison-Hooks}
\author[2,6]{Elijah Kane} 
\author[1]{Jared Lucey}
\author[3,4]{Philip Mauskopf}
\author[1]{Lynn Miles}
\author[1]{Sanetra Bailey Newman}
\author[1]{Gerard Quilligan}
\author[3]{Cody Roberson}
\author[1]{Salman Sheikh}
\author[1]{Manuel Vega}
\author[3]{Eric Weeks}
\author[1]{Christopher Wilson}
\author[1]{Travis Wise}
\affil[1]{NASA Goddard Space Flight Center, Greenbelt, MD, USA}
\affil[2]{NASA Jet Propulsion Laboratory, Pasadena, CA, USA}
\affil[3]{School of Earth and Space Exploration, Arizona State University, Tempe, AZ, USA}
\affil[4]{Department of Physics, Arizona State University, Tempe, AZ, USA}
\affil[5]{Department of Physics and Astronomy, Johns Hopkins University, Baltimore, MD, USA}
\affil[6]{Department of Physics, California Institute of Technology, Pasadena, CA, USA}

\authorinfo{Further author information: (Send correspondence to A.K.S.)\\A.K.S.: E-mail: asincla6@jh.edu}

\begin{document} 
\maketitle

\begin{abstract} 
We present further development and testing of prototype kinetic inductance detector (KID) readout electronics for the PRobe far-Infrared Mission for Astrophysics (PRIMA). PRIMA is a NASA Probe-class astrophysics mission concept with two science instruments, FIRESS and PRIMAger, which provide spectroscopic and polarimetric imaging from 24 to 261 um. Together these instruments will explore the far-infrared sky with spatial-spectral mapping speeds that exceed previous studies by orders of magnitude, opening a vast discovery space. This increase in mapping speed is achieved by densely packed ultra-sensitive KIDs and cryogenically cooled optics. The PRIMA flight readout electronics will use the build-to-print SpaceCube Mini v3.0 which houses a radiation-tolerant Kintex KU060 field programmable gate array (FPGA). A custom high-speed digitizer board is being designed along with an RF electronics (RFE) board. The RFE was designed to provide programmable gain, band filtering, and RF switching between instruments. The prototype readout electronics, including the RFE, use functionally equivalent parts to the flight parts. In this proceeding we measure the readout loopback noise performance in the second Nyquist zone and timestream cross-talk performance. We present preliminary sweeps with the prototype readout coupled to prototype FIRESS detectors.  Lastly, we demonstrate a cosmic ray pulse detection and real-time removal with an algorithm implemented in firmware.
\end{abstract}


\section{INTRODUCTION}
\label{sec:intro}  

The PRobe far-Infrared Mission for Astrophysics (PRIMA)\cite{Glenn2025} is a proposed NASA Astrophysics Probe Explorer with a planned launch date in 2032. The PRIMA mission concept is well timed for the far-infrared since the recent retiring of the Stratospheric Observatory for Infrared Astrophysics (SOFIA) and the demand for spectral continuity between James Webb Space telescope’s longest wavelengths and ALMAs shortest wavelengths. PRIMA is planning to fly an actively cooled 1.8 meter primary with two instruments, FIRESS and PRIMAger. FIRESS is a moderate resolution grating based imaging spectrometer spanning 24-235 um with an additional high resolution mode using a post-dispersed Fourier transform module. PRIMAger contains a hyperspectral imager and a four band polarimetric broad-band imager from 24-261 um. These instruments will enable spatial-spectral survey speeds that exceed previous studies by many orders of magnitude\cite{Bradford2025,Ciesla2025} mapping an enormous discovery volume. Specifically, PRIMA is going to help us understand the formation of planetary atmospheres, the buildup of dust and metals over cosmic time, and galactic evolution through studies of black hole accretion and feedback mechanisms. Many more community developed science cases envisioned for PRIMA have been documented in two volumes of a general observer program submitted to the arXiv \cite{GOvol12023,GOvol22025}.

Three main technologies are responsible for enabling these massive gains in sensitivity: The actively cooled optics, the adiabatic diamagentization refrigerator \cite{DiPirro2024}, and the Zodiacal background limited far-infrared sensitive kinetic inductance detectors (KIDs). Developed at JPL over 20 years ago \cite{Day2003}, KIDS have benefited from continued investment and have now become the leading far-Infrared detector technology that can be formed into large arrays and easily multiplexed. The readout for KIDs is accomplished with microwave frequency multiplexing electronics and cryogenic low noise amplifiers. The readout electronics have similarly benefited from a parallel effort of development and investment which has produced open-source systems that have been used at ground based observatories \cite{Golwala2012,McHugh2012} and flown on balloon-borne telescopes \cite{Gordon2016,Paiella2019}. The PRIMA readout electronics build on the heritage of these systems and in this proceeding we present the prototype developed for the phase A concept study. Overall architecture, component selection, and preliminary performance measurements have been previously reported \cite{Essinger-Hileman2026}. These proceedings present further laboratory measurements and architecture discussions, including system performance in the 2nd Nyquist zone used for the PRIMAger instrument, system crosstalk measurements, and results from measurements of a prototype FIRESS detector array. We also present measurements of a real-time cosmic ray glitch detection and removal algorithm.

\section{Phase A prototype readout electronics}
\label{electronics}

\subsection{Prototype Control Software Architecture}
The majority of the signal processing is done real-time in the FPGA gateware but certain parts of initialization, control, and calibration are accomplished within a soft-core processor implemented within the reprogrammable logic. Specifically, a MicroBlaze soft-core processor has been configured to run Petalinux with custom drivers and C applications. For instance bringing up the high-speed digitizers, which require JESD core configuration and SPI commanding, were given custom setup application programs. Another example is a detector calibration process which requires performing frequency sweeps across the band (see figure \ref{fig:firess_sweeps}). This process involves a C program working in concert with the real-time signal processing. 

External control is accomplished with a Raspberry Pi 4 through an Ethernet connection brought up by the MicroBlaze. The Raspberry Pi is a stand-in for the Command and Data Handling computer (C\&DH) to be delivered by the Italian Space Agency. The communication protocol will be switched from Ethernet to SpaceWire for the flight readout electronics.

\subsection{Second Nyquist Zone and Loopback Measurements}

\begin{table}[ht]
\caption{Readout frequency band definition for FIRESS and PRIMAger instruments with their Nyquist zone and sampling mode.} 
\label{tab:fonts}
\begin{center}       
\begin{tabular}{|l|c|c|c|} 
\hline
\rule[-1ex]{0pt}{3.5ex}  Instrument & Frequency Range & Nyquist Zone & Sampling Mode\\
\hline
\rule[-1ex]{0pt}{3.5ex}  FIRESS & 0.4 - 2.4 GHz & 1st & NRZ\\
\hline
\rule[-1ex]{0pt}{3.5ex}  PRIMAger & 2.6 - 4.9 GHz & 2nd & RF \\
\hline
\end{tabular}
\end{center}
\end{table} 

\begin{figure} [ht]
\begin{center}
\begin{tabular}{c c} 
\includegraphics[height=5cm]  {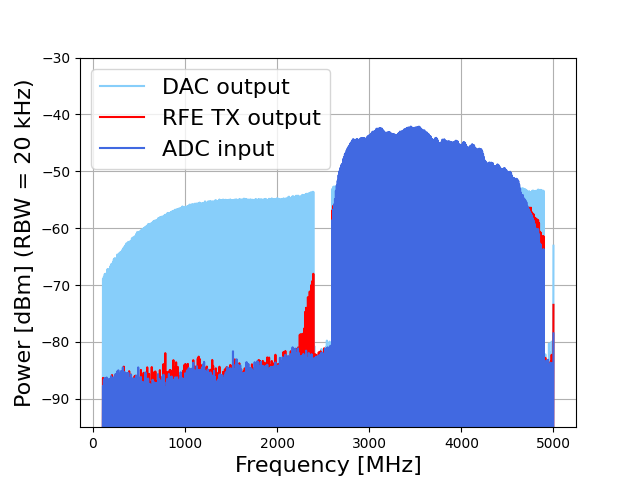} & \includegraphics[height=5cm]  {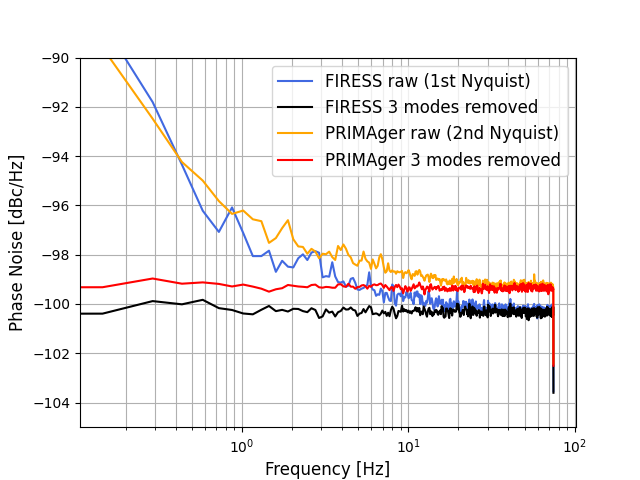}
\end{tabular}
\end{center}
\caption[example] { \label{fig:spectrum_and_median_psds} 
 At left is a spectrum analyzer measurement of a frequency comb generated for the PRIMAger band of 2.6 to 4.9 GHz. The output is measured at three different places along its path back to the ADC: directly at the output of the DAC in light blue, the output of the RF electronics transmit port in red, and right at the input port to the ADC in dark blue. At right is the median phase noise spectral density in dBc/Hz measured with 410 tones in RF loopback on the prototype's front panel while operating in the FIRESS mode and the PRIMAger mode. When removing the top three common modes from a principle component analysis the 1/f knee pushes below 0.1 Hz with a white noise level below -98 dBc/Hz.}
\end{figure}

The two instruments, FIRESS and PRIMAger, have different detector designs such that the readout resonant frequencies are in two different bands, FIRESS resonators span from 0.4 to 2.4 GHz and PRIMAger from 2.6 to 4.9 GHz. Only one instrument will be operated at a time which limits the required instantaneous bandwidth of the readout electronics to 2.5 GHz. Earlier kinetic inductance detector readouts used analog IQ mixers with tunable local oscillators to select the band centers \cite{McHugh2012, Gordon2016} but with advancements in high-speed digitizer technologies direct RF sampling has become an option. Eliminating analog IQ mixers reduces system complexity and cost along with associated non-linearity and imbalance challenges. Along with performance increases the high-speed digitizers provide alternative sampling modes from the standard non-return to zero (NRZ) which allow for different transfer functions. We chose to use the NRZ mode for FIRESS and sample in the 2nd Nyquist zone for PRIMAger using an RF D/A mode where halfway through a sample the output waveform is inverted. This RF D/A mode puts most of the power in the 2nd Nyquist zone as opposed to the NRZ mode which contains an unacceptably steep roll-off near 5 GHz. The left plot of figure \ref{fig:spectrum_and_median_psds} presents measurements of the spectrum of the prototype readout electronics when operated in this RF mode with a representative PRIMAger frequency comb. The spectrum was measured directly at the output of the D/A on the front panel then through the transmit path of the RFE which contains band defining filters and finally through both the transmit and receive paths of the RFE just before the A/D. 

Although operating in RF mode provides the correct bandwidth and center frequency the peak power is reduced when compared to operating in the first Nyquist zone with an NRZ sampling mode. When properly configured the D/A however is not the limiter in signal to noise for the readout system and is instead the A/D. To confirm this minimal impact on the readout noise we performed a loopback phase noise measurement in two configurations, FIRESS using NRZ mode and PRIMAger in RF mode. Figure \ref{fig:spectrum_and_median_psds} shows the median phase noise spectral density for both configurations running over 400 tones. The median phase noise performance in white noise and 1/f knees are within 1 dB which ensures the readout electronics will provide comparable performance for both instruments. The slightly higher median phase noise for the PRIMAger mode is expected in that the RF D/A sampling mode reduces the output power per tone by approximately 3 dB but also provides a substantially flatter response across the band of $\sim$1 dB when compared to the NRZ sampling mode slope of $\sim$3.9 dB which reduces the difference in medians.    

\subsection{Readout Electronics Cross-talk measurements}
\begin{figure} [ht]
\begin{center}
\begin{tabular}{c c} 
\includegraphics[height=5cm]  {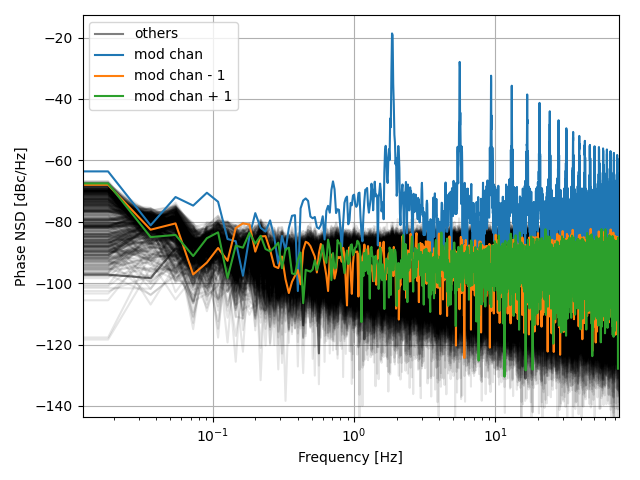} & \includegraphics[height=5cm]  {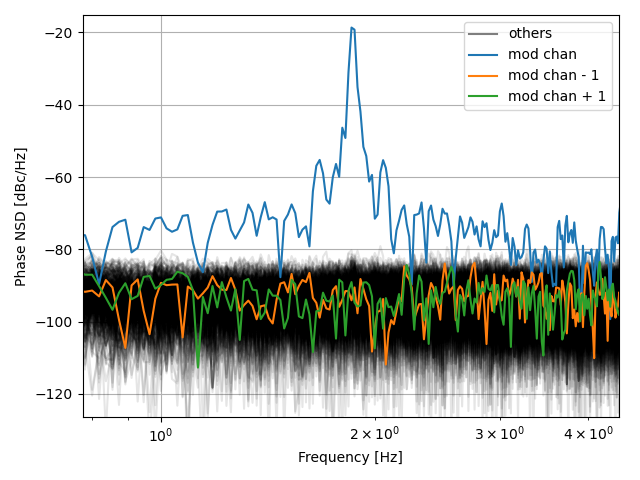}
\end{tabular}
\end{center}
\caption[example] { \label{fig:crosstalk} 
Phase noise spectral density in dBc/Hz measured with 512 tones in RF loopback on the prototype's front panel with a single channel phase modulated at approximately 1.8 Hz. With the randomly assigned tones spread across the band the difference between the modulated channel peak PSD and each other channel noise floor is at least 60 dB.}
\end{figure}

A key advantage of KIDs is the intrinsic frequency multiplexing which implies a shared communication channel for the detectors. Detector signal isolation while maintaining high multiplexing factors is achieved through careful resonator design and frequency placement. The relatively large instantaneous bandwidth of the detector arrays significantly reduces the resonator cross-talk occurring on the detector array. The readout electronics ideally would have a readout channel cross-talk below this intrinsic detector level which from Bisigello\cite{Bisigello2016} we can conservatively place at -40 dB. We achieve this low cross-talk by using a method common to the field of digital communications which is orthogonal frequency division multiplexing. We limit tone frequencies within the comb to multiples of the accumulated output data rate. This ensures that each tone is considered orthogonal to each other tone such that each channels frequency response contains null points at every other frequency. While mathematically there should be no cross-talk, exceedingly high clock jitter and sampling differences at the D/A and A/D can cause non-orthogonality and thus warrants measurement. To demonstrate this condition of orthogonality and low cross-talk, a loopback measurement was performed with a frequency comb where one channel was phase modulated with a square wave at approximately 1.8 Hz while simultaneously monitoring all other channels. 
Figure \ref{fig:crosstalk} shows the results of the measurement with a phase noise spectral density plot with all channels. The modulated channel is clearly visible above the noise floor of the other channels. A zoom in of the main modulation peak is shown on the right with the noise floor of all other channels at least 60 dB down. We also highlight neighboring frequency channels to the modulated channel in orange and green as mod chan-1/+1. The tone power was lowered for this test resulting in a higher median phase noise than what is achievable as shown in figure \ref{fig:spectrum_and_median_psds}. 

\section{Demonstrations with prototype FIRESS array}
A short wavelength (Band 1, nominally 24--43~\textmu m) FIRESS prototype array was delivered by JPL to Goddard Space Flight Center for testing. The detector array was packaged within a light tight aluminum housing without silicon lenses attached to the chip. The detector package was installed within a BlueFors dilution refrigerator and read out with two different cryogenic RF chains. One chain contains two Low Noise Factory (LNF) low noise amplifiers thermally controlled to 18 K and 70-100 K to match the estimated spacecraft temperature mounting points. The other cryogenic RF chain contains a low noise amplifier built by Arizona State University for 0.5 to 3 GHz and is mounted on the 4 K stage without active thermal control.

\begin{figure} [ht]
\begin{center}
\begin{tabular}{c c} 
\includegraphics[height=5cm]  {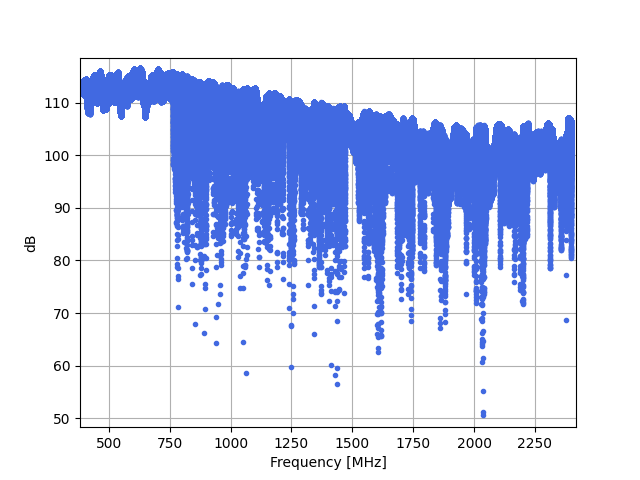} & \includegraphics[height=5cm]  {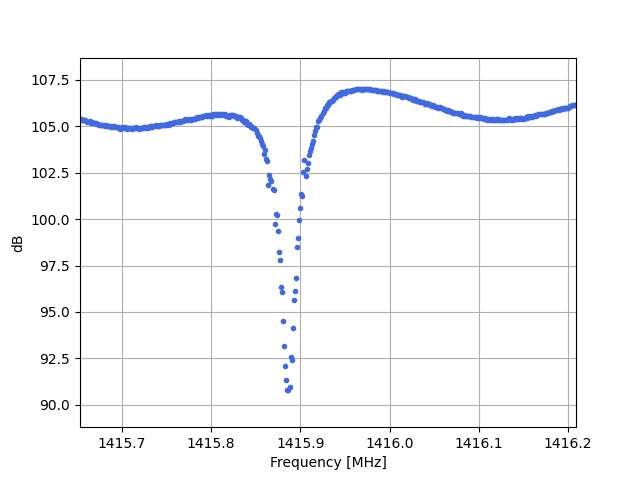}
\end{tabular}
\end{center}
\caption[example] { \label{fig:firess_sweeps} 
 Sweeps on a prototype FIRESS array at GSFC with the phase A prototype readout electronics. Left is a sweep of the FIRESS band with approximately 1 kHz frequency step resolution. The figure on the right is a zoom in on a randomly chosen resonator.}
\end{figure}

The first measurement presented in figure \ref{fig:firess_sweeps} is the latest frequency sweep of the FIRESS prototype array with the prototype readout electronics. Frequency sweeps are an important array characterization and calibration process that the readout is required to perform. This sweep used a comb for the FIRESS band and was performed with an application on the MicroBlaze which has a programmable frequency step size and span. In addition to the sweep functionality we are developing resonance finding and fitting functions which operate on the sweep output, pulling from the fields multiple open-source repositories for superconducting resonators.

\subsection{Real-time Cosmic Ray Glitch Removal}

PRIMA will observe from the Sun-Earth Lagrange point 2 (L2) which exposes the spacecraft to an increased flux of cosmic rays when compared to a low Earth orbit. The energy spectrum of particles is well known from previous telescopes at L2 \cite{Planck2014, Rauscher2025} and is used to carefully select components for the readout electronics. The detectors leverage decades of research around cosmic ray mitigation techniques for KIDs and have incorporated those principles into the design of both FIRESS and PRIMAger arrays\cite{Karatsu2019, Kane2024}. However, these techniques only lower the event rate on the detector and do not completely eliminate their effects. Therefore each detector is still expected to receive cosmic ray hits that will deposit energy. KIDs respond and dissipate the deposited energy quickly when compared to other detector technologies that have flown at L2. When a cosmic ray hits a detector it creates athermal phonons that  break Cooper pairs, rapidly changing the resonant frequency and creating a pulse with a fast rise time. Subsequent pair recombination and phonon dissipation lead to an exponentially decaying tail with characteristic times of less than 1~ms. The desired science data rate is in the range of 100-400 Hz which implies that multiple cosmic rays will hit the detectors over one science sampling period. Unmitigated the sample accumulation down to the final science rate would exhibit an elevated noise floor. This motivated the development of a real-time cosmic ray detection and removal algorithm. 

\begin{figure} [ht]
\begin{center}
\begin{tabular}{c c c} 
\includegraphics[height=6cm]  {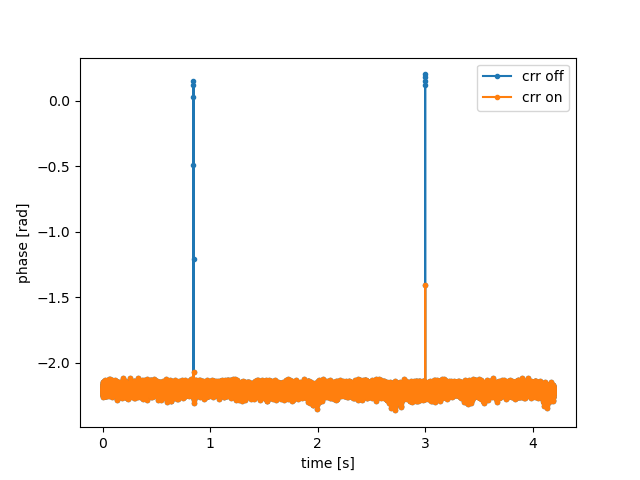} & \includegraphics[height=6cm]  {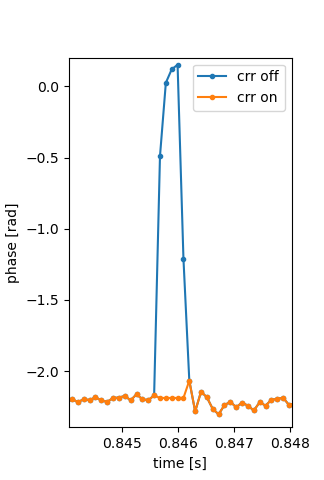} &
\includegraphics[height=6cm]  {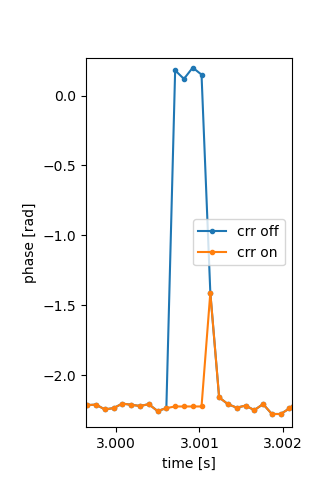}
\end{tabular}
\end{center}
\caption[example] { \label{fig:cosmic_ray_ts} A randomly chosen detector phase timestream from the FIRESS prototype array demonstrating the operation of the cosmic ray removal (CRR) module. The blue timestream on the left is the phase as a function of time with the CRR module disabled and the orange trace is the same detector channel with the CRR module enabled. The two cosmic ray pulses observed were detected and removed and replaced with last known good mean in real-time.
A zoom in on both pulses are shown on the right. For this preliminary test the de-trigger threshold was set higher than 5 standard deviations of the noise floor causing the second glitch to contain part of the falling edge of the pulse. This programmable de-trigger threshold can be lowered to 1 or 2 standard deviations. 
}
\end{figure}

Within the digital signal processing of the readout after digital down conversion and accumulation to a rate of approximately 9.5 kHz we have incorporated a cosmic ray pulse detection and removal module. The current algorithm carries a running mean which is compared to each new sample and if the absolute value of the difference between the mean and the new sample is greater than a  programmable high threshold then the system goes into the triggered state. When in the triggered state the module outputs the last known good mean and only de-triggers when the new value drops below another programmable low threshold. This method is low on resource utilization and performs well for the expected count rates. The module was implemented in prototype gateware and tested with the FIRESS prototype array to detect the ambient cosmic ray background within the lab at GSFC. Figure \ref{fig:cosmic_ray_ts} shows a 4 second timestream of a randomly chosen detector. The readout was configured to output the same channel twice but with one setup to detect and remove cosmic ray hits and the other in a pass-through mode. This allowed for direct comparison with and without glitch removal. The figure shows two timestreams where the blue trace is in the pass-through mode and the orange has cosmic ray detection and removal enabled. During the 4 second timestream two cosmic ray hits were observed and successfully removed by the module. For this test the de-trigger low threshold was set above 5-sigma of the noise level which causes the second pulse of figure \ref{fig:cosmic_ray_ts} to contain one point on the falling edge of the pulse. The programmable de-trigger threshold can be set much lower in future measurements. 
We are currently exploring multiple pulse detection and removal algorithms and we will test them with lab background cosmic rays as presented here and with a radioactive source in the near future. 

\section{Conclusion}
We have presented aspects of the architecture and measurements of the phase A detector readout electronics prototype not covered previously. We performed loopback noise measurements in both instrument modes showing that both have a similar noise level that meets the median noise requirements and 1/f. The electronics cross-talk between separate detector channels was measured to have at least 60 dB of isolation. The prototype hardware and software were used to perform preliminary sweeps across a prototype FIRESS detector array. The sweep was entirely performed by a C application on the MicroBlaze as intended in the flight configuration. Finally we implemented a cosmic ray glitch detection and removal module into firmware and demonstrated its real-time performance with the FIRESS prototype. We intend to continue developing control software on the MicroBlaze while using the readout for more laboratory measurements of the FIRESS array with far-infrared thermal loads and radioactive sources in the near future.


\bibliography{report} 
\bibliographystyle{spiebib} 

\end{document}